\documentclass{article}

\usepackage{arxiv}
\usepackage[utf8]{inputenc} 
\usepackage[T1]{fontenc}    
\usepackage{hyperref}       
\usepackage{url}            
\usepackage{booktabs}       
\usepackage{amsfonts}       
\usepackage{nicefrac}       
\usepackage{microtype}      
\usepackage{lipsum}		

\usepackage{graphicx}     
\usepackage{subcaption}
\usepackage{amsmath}
\usepackage{tikz}
\usetikzlibrary{matrix}
\usepackage{listings}

\title{On the Fibonacci sequence and the Linear Time Invariant Systems}

\author{
  J.M. Gorriz\thanks{gorriz@ugr.es, jg528@cam.ac.uk} \\
  Data Science and Computational Intelligence Institute\\
  University of Granada\\
  Granada, Spain \\
}

\begin{document}
\maketitle

\begin{abstract}
The Fibonacci sequence (FS) possesses exceptional mathematical properties that have captivated mathematicians, scientists, and artists across centuries. Its intriguing nature lies in its profound connection to the golden ratio, as well as its prevalence in the natural world, exhibited through phenomena such as spiral galaxies, plant seeds, the arrangement of petals, and branching structures. This report delves into the fundamental characteristics of the FS, explores its relationship with the golden ratio using Linear Time Invariant (LTI) systems, and investigates its diverse applications in various fields. Approaching the topic from the standpoint of a digital signal processing instructor in a grade course, we depict the FS as the consequential outcome of an LTI system when subjected to the unit impulse function. This LTI system can be regarded as the original source from which one of the most renowned formulas in mathematics emerges, and its parametric definition, along with the associated systems, is intricately tied to the golden ratio, symbolized by the irrational number $\Phi$. This perspective naturally elucidates the well-established intricate relationship between the FS and $\Phi$. Furthermore, building upon this perspective, we showcase other LTI systems that exhibit the same magnitude in the frequency domain. These systems are characterized by either an impulse response or a difference equation, resulting in a comparable or ``equivalent'' FS in terms of absolute value. By exploring these connections, we shed light on the remarkable similarities and variations that arise within the FS under different LTI systems.
\end{abstract}


\section{Introduction}

The Fibonacci sequence (FS) is a series of numbers in which each number is the sum of the two preceding ones, typically starting with 0 and 1 \cite{Vries16}. Thus, the sequence begins as follows: 
\begin{equation}\label{eq:fibonacci}
\{0,1,1,2,3,5,8,13,21,34,55,\ldots,\}     
\end{equation}
The FS was first introduced to the Western world by Leonardo Fibonacci, an Italian mathematician, in his book "Liber Abaci" in 1202 \cite{Sigler02}. The FS is relevant and significant in several areas:
\begin{itemize}
    \item Mathematics: It appears in various mathematical concepts, including number theory, combinatorics, and geometry. The sequence is often used as an example in teaching mathematical concepts and problem-solving techniques \cite{Vries16}.
    \item Nature: It can be observed in numerous natural phenomena, such as the growth patterns of plants, the arrangement of leaves on stems, the branching of trees, the spirals of shells, storm systems like hurricanes and tornadoes and the distribution of seeds in a sunflower. These occurrences demonstrate the presence of the FS and its connection to the principles of growth and optimization in living organisms \cite{Grigas13}.
    \item Art and Design: It is prevalent in art, design, and aesthetics \cite{Power19}. Artists, architects, and designers often incorporate these principles in their work to create visually pleasing compositions, harmonious proportions, and balanced structures. The golden ratio is believed to contribute to the perceived beauty and balance in artwork and design.
    \item Financial Markets: It has been applied to financial analysis and trading \cite{Merrill77}. Traders and analysts use Fibonacci retracement levels and extensions as technical indicators to identify potential support and resistance levels in financial markets. These levels are based on ratios derived from the FS and are believed to help predict price movements.
    \item Computer Science: It is often used as an example for recursive functions, dynamic programming, and algorithmic analysis \cite{Abramovich19}. The sequence's recursive nature makes it useful for understanding and solving problems in computer programming and algorithm design.
\end{itemize}

Overall, the FS is relevant across multiple disciplines, highlighting its intrinsic mathematical beauty and its manifestation in the natural world. Its connections to nature, art, design, finance, and computer science make it a fascinating and versatile concept that continues to inspire research, exploration, and practical applications \cite{Brasch12}.

The FS exhibits a unique mathematical property known as the golden ratio (GR) $\Phi$ \cite{Livio02}, an irrational number equal to $\Phi = (1 + \sqrt{5} )/ 2$. The ratio between consecutive Fibonacci numbers tends to approach the golden ratio as the sequence approaches infinity. Specifically, dividing a Fibonacci number by its predecessor or successor yields an approximation of the golden ratio. It is considered aesthetically pleasing and has been utilized in various artistic and design principles. The ratio is believed to create harmonious proportions that are visually appealing to humans. 

This report showcases the description of the Fourier Series (FS) as the impulse response of a causal, non-stable Linear Time Invariant (LTI) system within the realm of digital signal processing theory. By adopting this innovative approach, we are able to attribute similar properties exhibited by various systems, which are commonly defined using difference equations, to the FS.

\begin{figure}
\centering
\begin{tikzpicture}
\matrix (a)[row sep=0mm, column sep=0mm, inner sep=1mm,  matrix of nodes] at (0,0) {
\includegraphics[width=\textwidth]{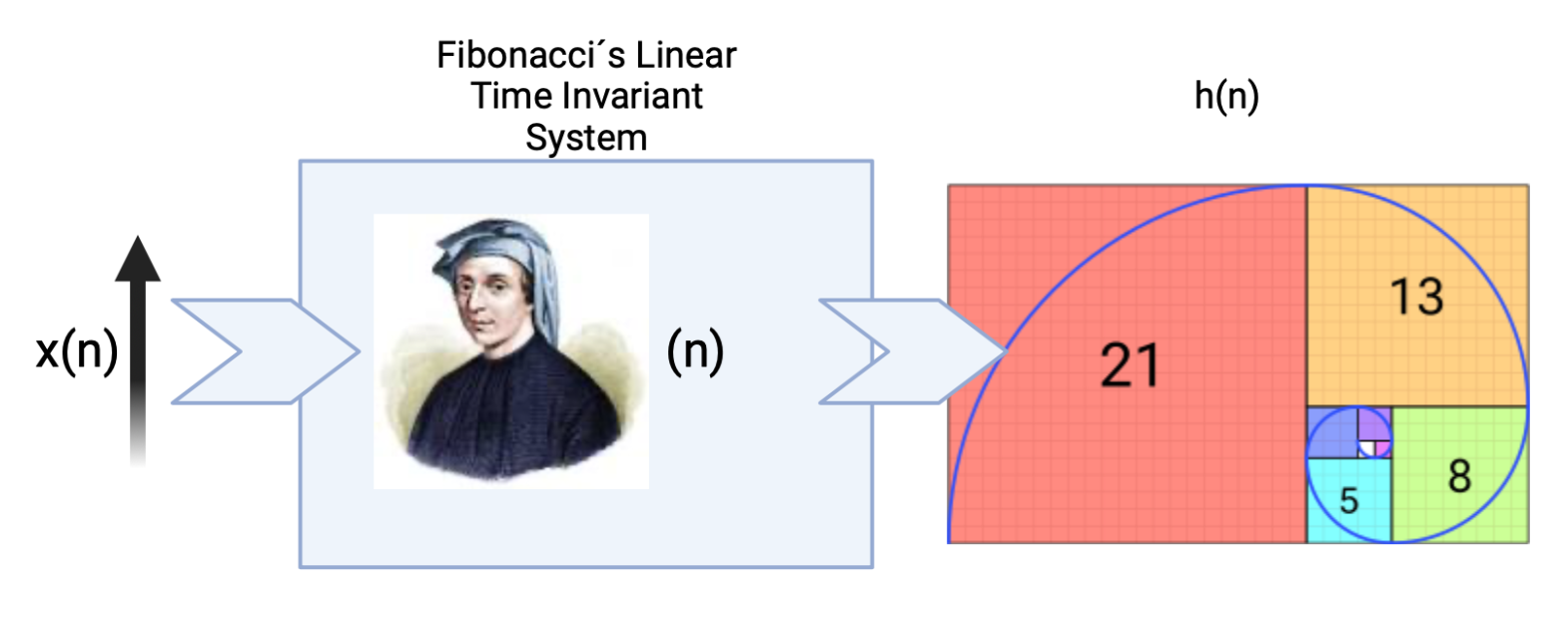}\\\\
        };
\end{tikzpicture}
\caption{The impulse response of the FS system, the FS sequence.}
\label{fig:freqz}
\end{figure}

\section{Definition and properties}

The Fibonacci sequence (FS) finds its definition through various perspectives in geometry and mathematics \cite{Ferguson66,Flores67,Gabai70,Kalman82,Kessler04,Lee01,Levesque85,Miles60,Dresden09}. It represents a sequence of integers generated by the addition of the two preceding numbers, as demonstrated in Equation \ref{eq:fibonacci}. The analytical definition of this sequence can be succinctly expressed by considering the relationship between the elements $n$, $n-1$, and $n-2$ of the series:
\begin{equation}\label{eq:dos}
f_n=f_{n-1}+f_{n-2};\quad n>2
\end{equation}
or by the famous Euler-Binet's (or Moivre's) formula:
\begin{equation}\label{eq:dosbis}
f_n=\frac{\Phi^n-(-\Phi)^{-n}}{\sqrt{5}}
\end{equation}
The latter equation can be viewed as an outcome resulting from a direct property of the golden ratio: $\Phi^2=\Phi+1$. This property is also satisfied by its negative inverse $-\Phi^{-1}$, and by evaluating it in a recursive manner, we can express it as follows: "
\begin{equation}\label{eq:GRrec}
\Phi^n=f_n\Phi +f_{n-1};\quad \left(-\Phi^{-1}\right)^n=f_n\left(-\Phi^{-1}\right)+f_{n-1}
\end{equation}
By subtracting both expressions we obtain equation \ref{eq:dos}. Some other interesting properties of the FS are \cite{Grigas13}:
\begin{itemize}
\item This sequence satisfies the following property, where the golden ratio $\Phi$ naturally arises \cite{Kepler66}:
\begin{equation}\label{prop1}
\lim_{n\xrightarrow{}\infty}\frac{f_{n+1}}{f_{n}}=\Phi;
\end{equation}
\item Divisibility: Some interesting divisibility properties exist within the FS, such as the fact that any two consecutive Fibonacci numbers are relatively prime.
\item $f_n$ is a Fibonacci number if and only if $5f_n^2+4$ or $5f_n^2-4$ is a square number.

\item Given $f_n$ then we can compute $f_{n+1}=round(f_n\Phi)$

\item The odd and even indexed numbers of the FS can be determined by previous FS numbers \cite{Garland87}, such as: $f_{2n-1} = f_{n-1}^2 + f_n^2$; $f_{2n} = (2f_{n-1}+f_n)f_n$.

\item Etc.

\end{itemize}

The k-generalized Fibonacci numbers, including the Tribonaccis, Tetranaccis, and others, can be incorporated into the FS framework. This can be achieved by utilizing expressions similar to equation \ref{eq:dosbis}, which deviate from difference equations as discussed in Dresden's work \cite{Dresden09}. In the upcoming sections, we will examine the FS from an alternative standpoint rooted in digital signal processing theory. Central to this theory are the concepts of Linear Time Invariant (LIT) systems and the Z transform. These powerful ideas and tools facilitate the comprehension and resolution of recursive time series.

\section{The Linear Time Invariant system of the FS}

The Fibonacci sequence can be readily characterized as the impulse response of a LTI system. Specifically, this LTI system is an Infinite Impulse Response (IIR) system, which is represented by a difference equation as follows:
\begin{equation}\label{ec:diff}
y(n)=y(n-1)+y(n-2)+x(n)
\end{equation}
where $y(n)$ is the output of the system, $x(n)$ is the input and $n$ refers to the discrete-time domain.  With this in mind, and replacing $x(n)$ by Kronecker's function $\delta(n)$ in the time difference equation we readily see that the impulse response of the LTI system is the FS:
\begin{equation}
h(n)=h(n-1)+h(n-2)+\delta(n)
\end{equation}
This system can be analysed using the Z transform as the following. Assuming the aforementioned system to be causal, the region of convergence (ROC: from the farthest pole to infinite) is linked to the pole positions of the transfer function $H(z)$ ,where $z$ is a complex variable in the Z domain:
\begin{equation}
H(z)=\frac{Y(z)}{X(z)}=\frac{1}{1-z^{-1}-z^{-2}}=\frac{1}{(1-\Phi z^{-1})(1-\Tilde{\Phi}z^{-1})}
\end{equation}
where the poles of the system are placed in irrational positions $\Phi$ and $\Tilde{\Phi}=-\Phi^{-1}$. (Not) Surprisingly the roots of the denominator are $p=\frac{1\pm\sqrt{5}}{2}$, that is, the GR and other irrational number whose module is less than 1, $\Tilde{\Phi}=\frac{1-\sqrt{5}}{2}=-\frac{1}{\Phi}$ that shares several properties with the GR. In the Fourier domain, the described system can be represented as a symmetrical stop-band filter with characteristic frequencies located at $w_1=0.2\pi$ and $w_2=0.8\pi$ radians per sample (refer to Figure \ref{fig:freqz}).

\begin{figure}
\centering
\begin{tikzpicture}
\matrix (a)[row sep=0mm, column sep=0mm, inner sep=1mm,  matrix of nodes] at (0,0) {
\includegraphics[width=\textwidth]{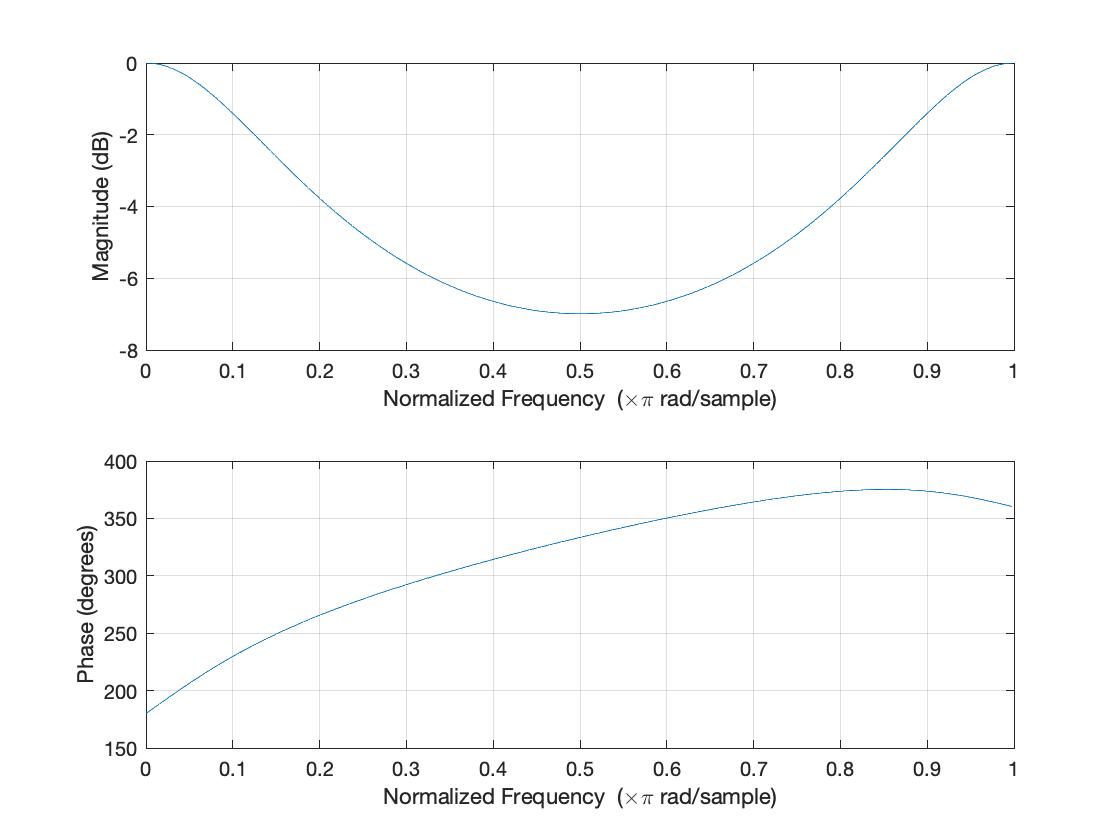}\\\\
        };
\end{tikzpicture}
\caption{The FS system in the Fourier domain.}
\label{fig:freqz}
\end{figure}

\begin{figure}
\centering
\begin{tikzpicture}
\matrix (a)[row sep=0mm, column sep=0mm, inner sep=1mm,  matrix of nodes] at (0,0) {
\includegraphics[width=\textwidth]{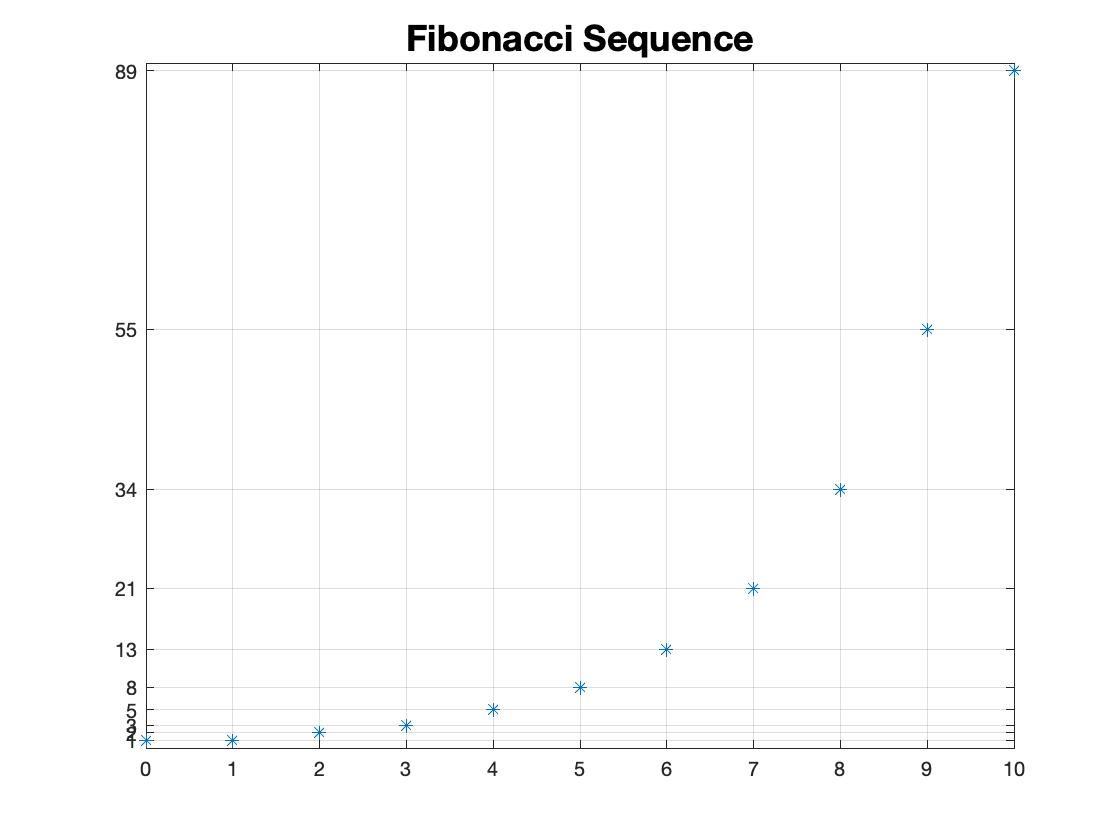}\\\\
        };
\end{tikzpicture}
\caption{The Fibonacci Sequence obtained as the impulse response of a LTI system.}
\label{fig:fibonacci}
\end{figure}

Computing the inverse Z transform of the transfer function we obtain:
\begin{equation}\label{eq:imp}
h(n)=C\Phi^n u(n)+ D\Tilde{\Phi}^nu(n)=\frac{1}{\sqrt{5}}(\Phi^{n+1}-\Tilde{\Phi}^{n+1})u(n)
\end{equation}
where $u(n)$ is the step function and $C=\frac{1}{1-\Tilde{\Phi}\Phi^{-1}}=\frac{\Phi+1}{\Phi+2}=\frac{\Phi}{\sqrt{5}}$ and $D=\frac{1}{1-\Phi\Tilde{\Phi}^{-1}}=\frac{1}{\Phi+2}=-\frac{\Tilde{\Phi}}{\sqrt{5}}$. The impulse response function of the LTI system is a new presentation of the Euler-Binet's formula and can be used to generate any element of the FS without using the recursion equation (see figure \ref{fig:fibonacci}). As an example, for the Ramanujan's magic number $n=1729$, the FS element is $h(1789)=\frac{1}{\sqrt{5}}(\Phi^{1789}-\Tilde{\Phi}^{1789})$. Notably, it should be emphasized that the initial non-zero element of the sequence in $h(n)$ is not $n=2$ as stated in equation \ref{eq:dos} with initial conditions $f_0=0$ and $f_1=1$. Instead, it occurs at $n=0$. Consequently, we can evaluate this sequence in a similar manner to the Euler-Binet's formula.

\subsection{Non-causal extension of the FS LIT system}

\begin{figure}
\centering
\begin{tikzpicture}
\matrix (a)[row sep=0mm, column sep=0mm, inner sep=1mm,  matrix of nodes] at (0,0) {
\includegraphics[width=\textwidth]{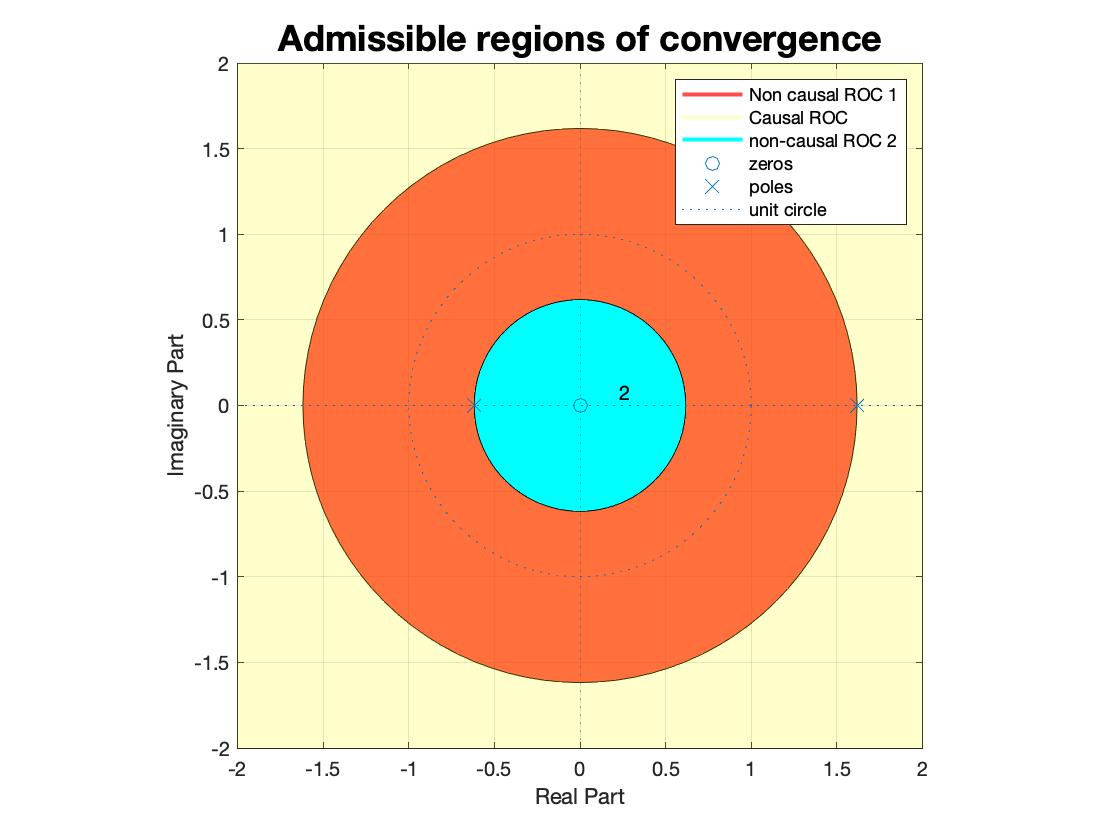}\\\\
        };
\end{tikzpicture}
\caption{All admissible ROCs of the LTI system $H(z)=\frac{1}{1-z^{-1}-z^{-2}}$.}
\label{fig:ROC}
\end{figure}

As evident from the derivation presented earlier, we made the assumption that the linear time-invariant (LTI) system under consideration was causal. Consequently, the series representing the impulse response was determined analytically through equation \ref{eq:imp}. In this analysis, we assumed that the region of convergence (ROC) for the system was given by $|z|>\Phi$ (refer to figure \ref{fig:ROC}). While this causal sequence is clearly observed in natural systems, it raises the question of what other sequences can be obtained from this LTI system if we disregard the assumption of causality inherent in the FS difference equation. The subsequent figure illustrates two possibilities: $\Tilde{\Phi}<z<\Phi$ and $\Tilde{\Phi}>z$. Consequently, the resulting impulse responses are as follows:

\begin{equation}\label{eq:impnc}
\begin{array}{c}
h(n)=-C\Phi^n u(-n-1)+ D\Tilde{\Phi}^nu(n)  \\
h(n)=-C\Phi^n u(-n-1) -D\Tilde{\Phi}^nu(-n-1)
\end{array}
\\
\end{equation}

\begin{figure}
\centering
\begin{tikzpicture}
\matrix (a)[row sep=0mm, column sep=0mm, inner sep=1mm,  matrix of nodes] at (0,0) {
\includegraphics[width=0.49\textwidth]{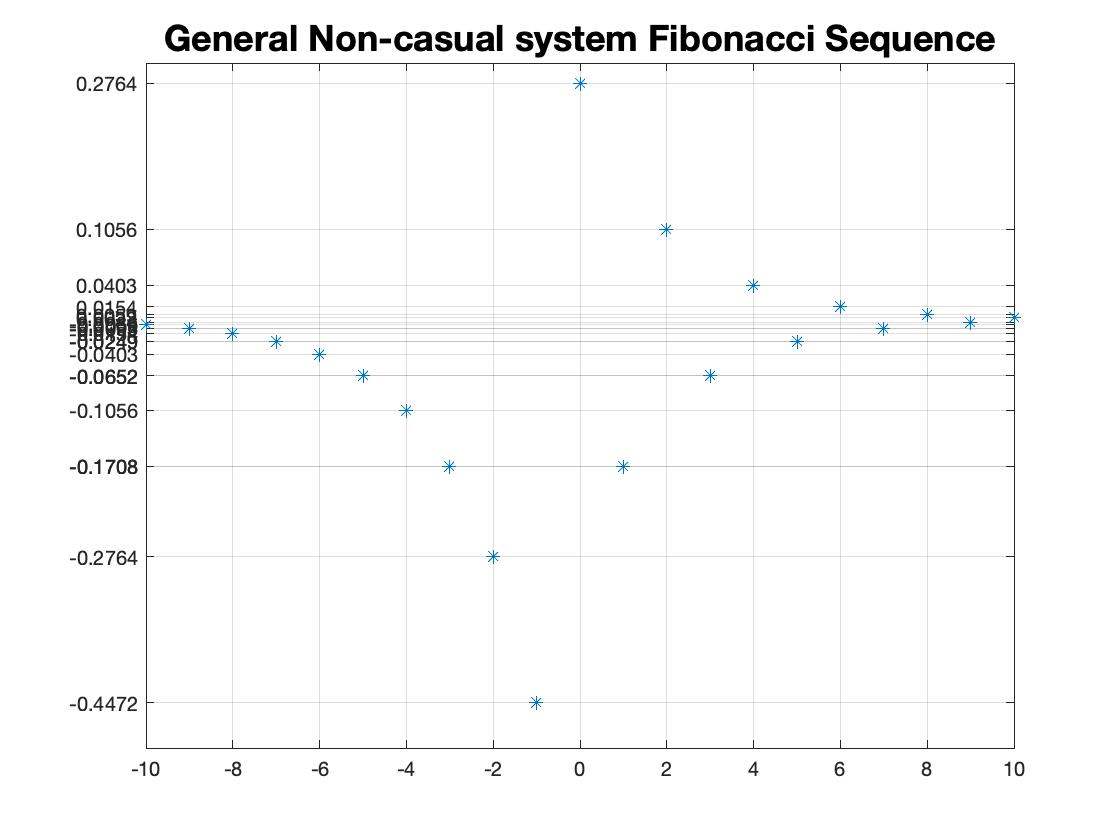}
\includegraphics[width=0.49\textwidth]{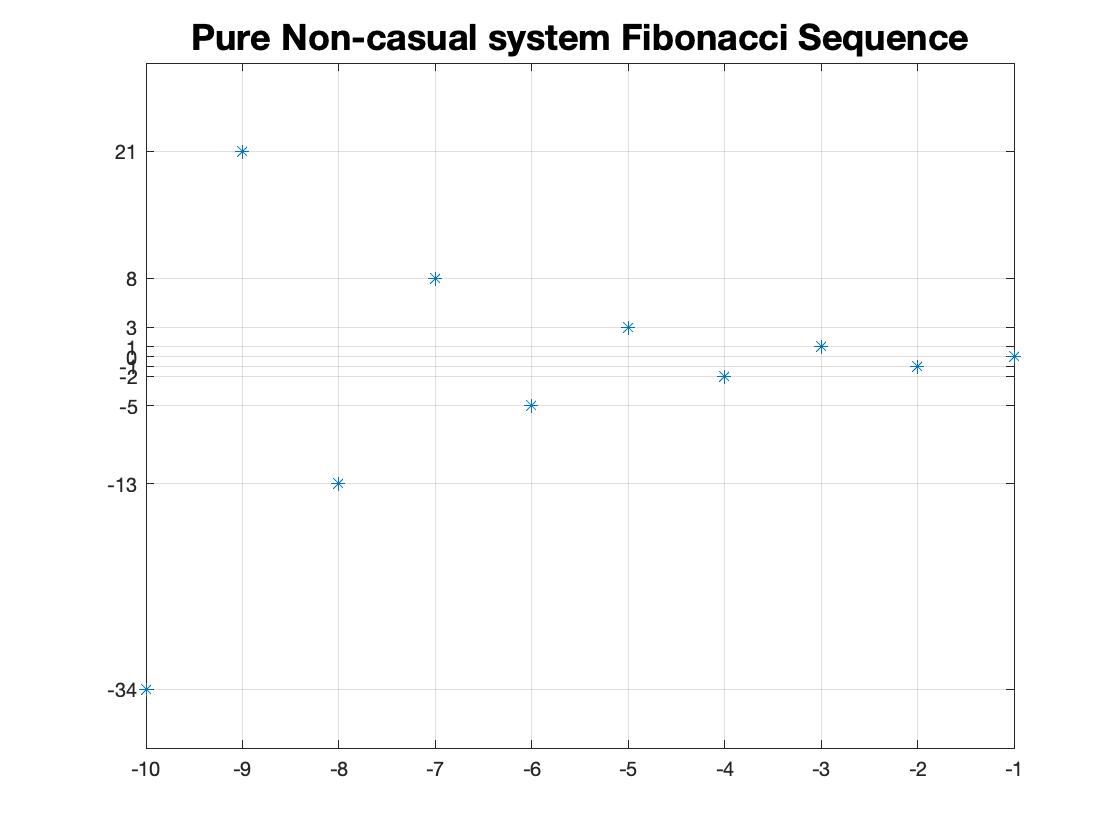}\\\\
        };
\end{tikzpicture}
\caption{The FS obtained as the impulse response of a non-causal LTI system .}
\label{fig:fibonacci2}
\end{figure}

The figure \ref{fig:fibonacci2} illustrates the plotted extended FS sequences, which represent impulse responses from non-causal LTI systems. Notably, the pure non-causal system ROC (cyan area in figure \ref{fig:ROC}) yields a sequence of integers reminiscent of the original FS. However, this sequence begins from zero in reverse order and alternates signs:
\begin{equation}
\{\ldots, 55, -34, 21, -13, 8, -5, 3, -2, 1, -1, 0\},  
\end{equation}
Thus, the FS sequence can be obtained by taking the absolute value of the output of this system and reversing the sequence.

\subsection{Is really the FS LIT system special?}

After careful analysis, it has been determined that the system under consideration is a mixed-linear phase IIR system with causal behavior. Specifically, the system possesses one pole within the unit circle denoted by $\Tilde{\Phi}$ and another pole outside the circle denoted by $\Phi$. Now, let's examine the LIT system, which exhibits the same magnitude with $\Tilde{H}(z)=H(z^{-1})$. In other words, the poles of the LIT system can be found in the set ${\Phi^{-1},-\Phi}$. As a result, for values of $n$ greater than zero, the difference equation governing the impulse response can be expressed as follows:
\begin{equation}
\Tilde{h}(n)= \Tilde{h}(n-2)-\Tilde{h}(n-1)
\end{equation}
with a sequence that is equivalent to:
\begin{equation}
\{1,-1,2,-3,5,-8,13,-21,34,-55,\ldots,\}   
\end{equation}
Hence, the FS can be derived from this impulse response by directly obtaining the absolute value of the sequence.

\begin{figure}
\centering
\begin{tikzpicture}
\matrix (a)[row sep=0mm, column sep=0mm, inner sep=1mm,  matrix of nodes] at (0,0) {
\includegraphics[width=\textwidth]{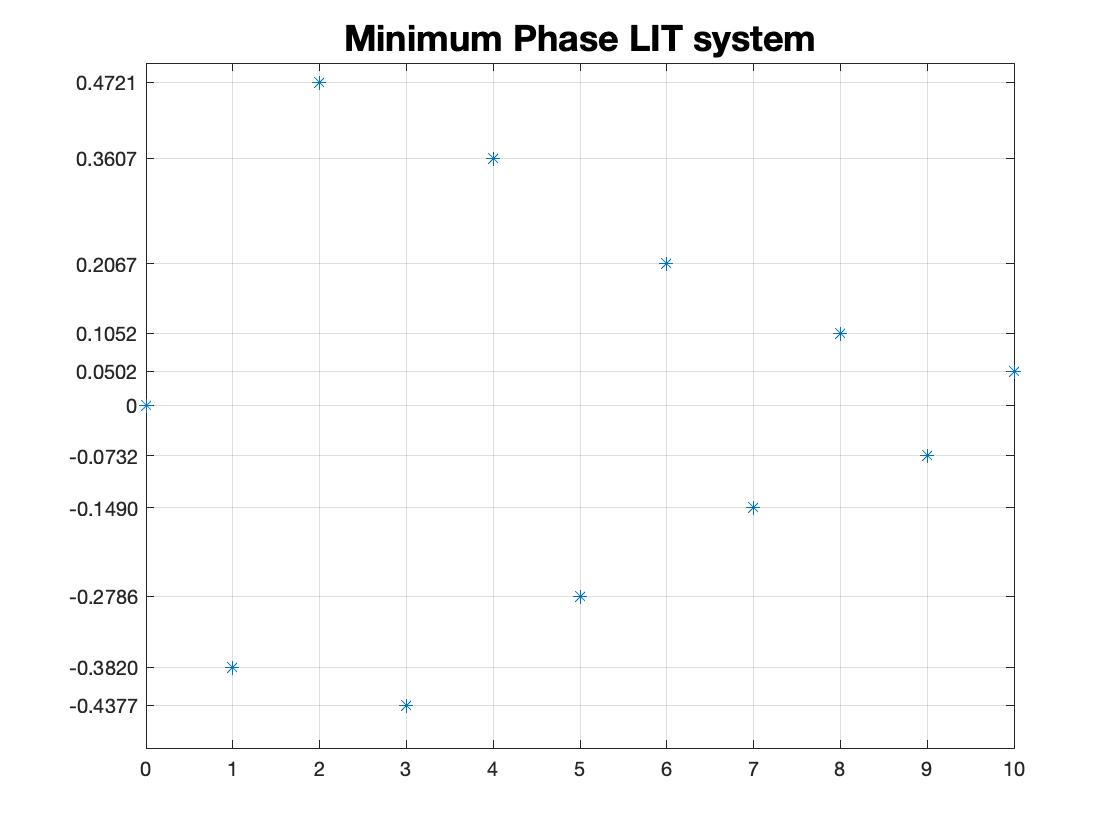}\\\\
        };
\end{tikzpicture}
\caption{Impulse response of the Minimum Phase LTI system $\hat{H}(z)$.}
\label{fig:MinimumPhase}
\end{figure}

One potential approach involves examining the causal minimum phase LIT system of the FS system after reversing the pole located outside the unit circle in $H(z)$. This newly introduced system, denoted as $\hat{H}(z)=\frac{z^{-1}\Tilde{\Phi}}{(1-\Phi^{-1}z^{-1})^2}$, can be represented by a distinct difference equation:
\begin{equation}
y(n)=\Tilde{\Phi}x(n-1)+2\Phi^{-1}y(n-1)-\Phi^{-2}y(n-2)
\end{equation}
and impulse response:
\begin{equation}
h(n)=-n\Phi^{-n}u(n)
\end{equation}
This causal and stable phenomenon exhibits a sequence of irrational numbers, leading to a convergence towards zero. The convergence is visually demonstrated in Figure \ref{fig:MinimumPhase}.

\section{Consequences of the proposed model}

The impulse response in the theory of LTI systems completely defines the behavior of any system. By convolving an input signal $x(n)$ with the impulse response $h(n)$, we can determine the corresponding output signal as $y(n) = x(n) * h(n)$. Consequently, we can analyze the possible outputs resulting from modifying the input of the FS LTI system, which is characterized by its corresponding IIR. This interpretation leads to several significant consequences, which are summarized in this section. For instance, a well-known relationship involving $\Phi$, as represented by equation \ref{prop1}, can be easily demonstrated using the difference equation presented in equation \ref{ec:diff}.

\subsection{The response of the FS LTI system to any input sequence}
The expression for a sequence in discrete time, denoted as $x(n)$, can be represented as an infinite sum of weighted impulse functions: $x(n) = \sum_{k=-\infty}^{\infty} x(k) \delta(n-k)$. When considering a causal, FS LTI system, the resulting output is:
\begin{equation}
y(n)=\sum_{k=-\infty}^{\infty}x(k)h(n-k)=\frac{1}{\sqrt{5}}\sum_{k=-\infty}^{n}x(k)(\Phi^{n-k+1}-\Tilde{\Phi}^{n-k+1})
\end{equation}
Using the property in equation \ref{eq:GRrec} it yields:
\begin{equation}
y(n)=\sum_{k=-\infty}^{n}x(k)f_{n+1-k}
\end{equation}
Therefore, it follows as anticipated that $h(n)$ equals $f_{n+1}$.

\subsection{A train of impulses and the step function response}

Given a train of shifted impulses $x(n)=\sum_{m=0}^{M-1}\delta(n-m)$ the output of the FS LTI system is:
\begin{equation}
y(n)=\sum_{m=0}^{M-1}h(n-m)=\sum_{m=0}^{M-1}f_{n+1-m}
\end{equation}

In general, the output of the FS LTI system with input $x(n)=u(n)$ in the Z domain is given by:
\begin{equation}\label{eq:step0}
Y(z)=H(z)X(z)=\frac{1}{1-z^{-1}-z^{-2}}\frac{1}{1-z^{-1}}
\end{equation}
Thus, its inverse Z transform yields:
\begin{equation}\label{eq:step}
y(n)=A\Phi^n u(n)+ B\Tilde{\Phi}^nu(n)+Cu(n)
\end{equation}
where $A=\frac{1}{(1-\Phi^{-1}\Tilde{\Phi})(1-\Phi^{-1})}=\frac{2\Phi+1}{2\Phi-1}$, $B=\frac{1}{(1-\Tilde{\Phi}^{-1}\Phi)(1-\Tilde{\Phi}^{-1})}=\frac{1}{4\Phi+3}
$ and $C=\frac{1}{(1-\Phi)(1-\Tilde{\Phi})}=-1$. On the other hand, in the time domain we can write:
\begin{equation}
y(n)=\sum_{k=-\infty}^{n}u(k)f_{n+1-k}=\sum_{k=0}^{n}f_{n+1-k}=f_{n+3}-1
\end{equation}

\begin{figure}
\centering
\begin{tikzpicture}
\matrix (a)[row sep=0mm, column sep=0mm, inner sep=1mm,  matrix of nodes] at (0,0) {
\includegraphics[width=\textwidth]{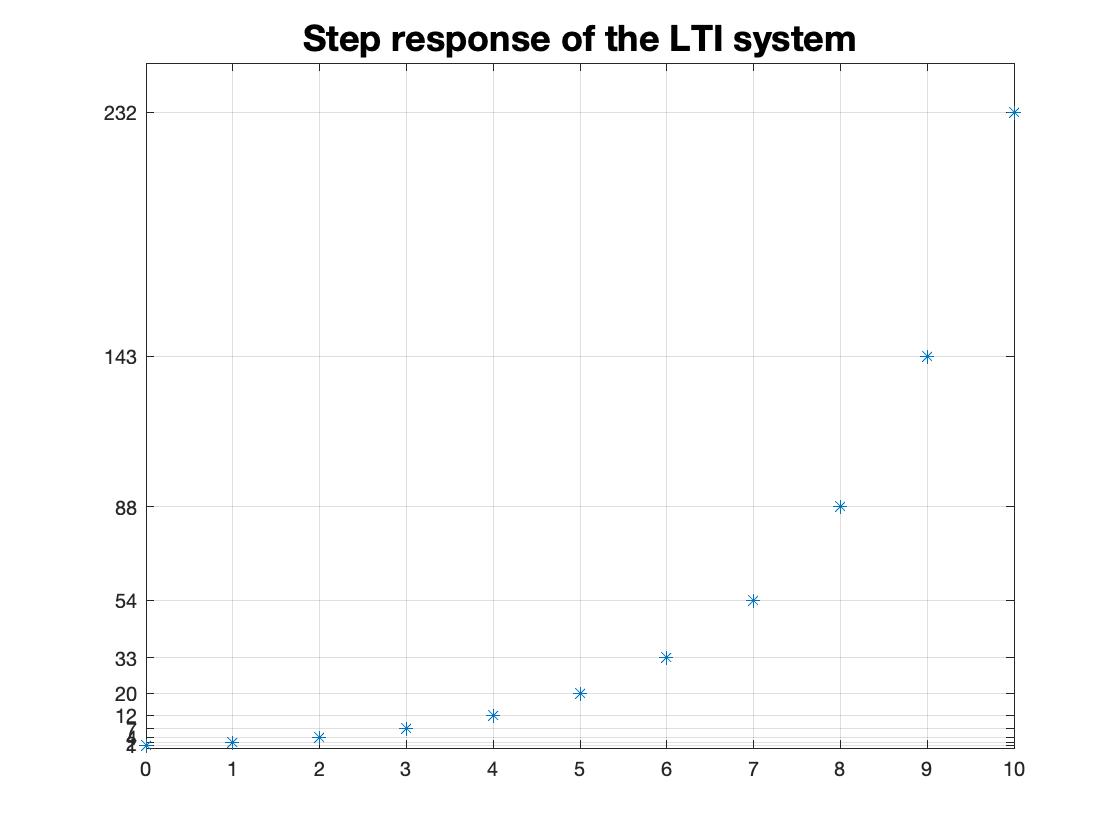}\\\\
        };
\end{tikzpicture}
\caption{The Fibonacci Sequence obtained as the step function response of a LTI system.}
\label{fig:step}
\end{figure}

In figure \ref{fig:step} we plotted the sequence:
\begin{equation}
\{1,2,4,7,12,20,33,54,88,\ldots\}     
\end{equation}
which is the Fibonacci meanders sequence, that is, the FS minus one \cite{Sloane73}.

\section{Conclusions}

The FS sequence is commonly believed to be the origin of various natural phenomena. Rather than being solely a source, it can be regarded as an outcome of a unique causal linear time-invariant (LTI) system, often referred to as the ``engine'' which generates it along with its distinct characteristics. In this report, we present evidence that the poles situated at $\Phi$ and $\Tilde{\Phi}$ determine the impulse response of an LTI system, leading to the FS. This system can be expanded to encompass non-causal, minimum phase systems that exhibit similar properties to the original one. The resulting output sequences are closely linked to the renowned Fibonacci sequence and several others, achieved by employing different inputs for the proposed systems.

\section*{Acknowledgments}
Thanks to my students at the University of Granada.

\appendix

\section{Some representations of the FS:}

The FS can be obtained from the impulse response of a LTI system as shown in previous sections:
\begin{equation}\label{eq:ap1}
h(n)=D[\Phi^{n+2}+ \Tilde{\Phi}^{n}]=D[\Phi^{n+2}+ (-1)^n\Phi^{-n}];\quad n\geq 0
\end{equation}
where $D=\frac{1}{1+\Phi^2}$. For $n\geq 0$ we have :
\begin{equation}\label{eq:ap2}
\lim_{n\xrightarrow{}\infty}\frac{h(n)}{h(n-1)}=\frac{[\Phi^{n+2}+ (-1)^n\Phi^{-n}]}{[\Phi^{n+1}+ (-1)^{n-1}\Phi^{-n+1}]}=\Phi
\end{equation}
Some closed-form expressions \cite{Ball03} such as the Binet's formula:
\begin{equation}\label{eq:ap3}
f(n)=\frac{1}{\sqrt{5}}(\Phi^n-(-\Phi)^{-n}); \quad n\geq 1
\end{equation}
can be rewritten for $n \geq 0$ as:
\begin{equation}\label{eq:ap4}
f(n)=\frac{1}{\sqrt{5}}(\Phi^{n+1}-(-\Phi)^{-n-1}); \quad n\geq 0
\end{equation}
From equation \ref{eq:ap1} we can write:
\begin{equation}\label{eq:ap5}
h(n)=\frac{\Phi^{n}}{1+\Tilde{\Phi}^{2}}+ \frac{\Tilde{\Phi}^{n}}{1+\Phi^{2}};\quad n\geq 0
\end{equation}
Then, after some manipulations and using $\Phi^2=\Phi+1$ we have:
\begin{equation}\label{eq:ap6}
h(n)=\frac{\Phi^n(\Phi^2+1)+\Tilde{\Phi}^n(\Tilde{\Phi}^2+1)}{5}=\frac{1}{\sqrt{5}}(\Phi^n\Phi+\Tilde{\Phi}^n\Tilde{\Phi});\quad n\geq 0
\end{equation}
that equals to \ref{eq:ap4}.

\section{Some exercises in Matlab:}

In this report, all the findings can be effectively represented using Matlab. Specifically, the following functions prove to be valuable for representing the various aspects discussed:
\begin{itemize}
\item roots: This function calculates the roots and poles of the system function $H(z)$.
\item freqz: It provides the frequency response of a systems function $H(z)$.
\item conv: This function performs the multiplication of two polynomials.
\item impz: It enables the computation of the impulse response of the system function $H(z)$.
\end{itemize}

\begin{figure}
\centering
\begin{tikzpicture}
\matrix (a)[row sep=0mm, column sep=0mm, inner sep=1mm,  matrix of nodes] at (0,0) {
\includegraphics[width=\textwidth]{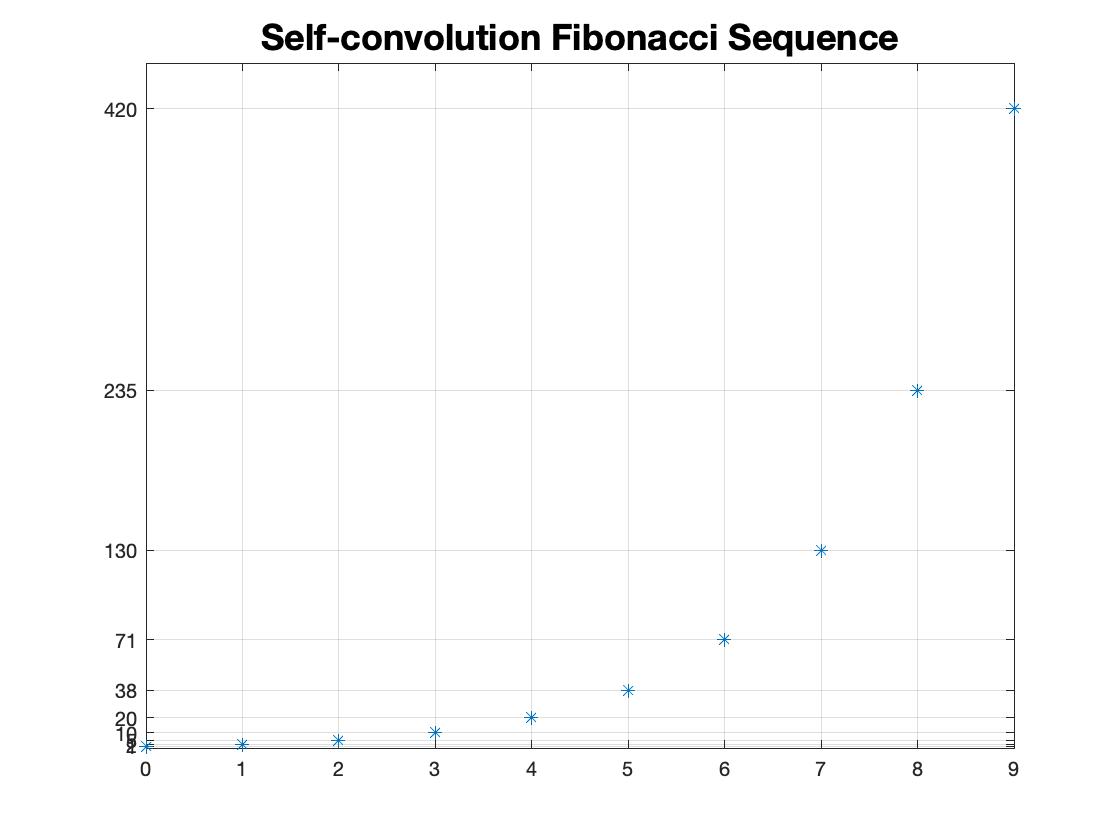}\\\\
        };
\end{tikzpicture}
\caption{The Self convolution FS}
\label{fig:doubleFS}
\end{figure}

For instance, let's consider an example where we want to compute the impulse response of the self-convolution of Fibonacci numbers. To obtain this sequence, we introduce the FS into a FS system, either by combining two systems in series or using a "convolution" FS system:
\begin{equation}\label{eq:ap7}
H_{conv}(z)=H(z)H(z) \leftrightarrow h_{conv}(n)=h(n)*h(n)
\end{equation}
The following code gives the required series:
\begin{lstlisting}[language=Matlab]
B=1;   % The numerator of the FS  system
A=[1 -1 -1];   % The denominator of the FS system
figure
freqz(B,A);   % The frequency response of the FS system
Aconv=conv(A,A);   % Multiplying both polynomials
figure
freqz(B,Aconv);   % Frequency response of the conv FS system
[h,t] = impz(B,Aconv,10); % Impulse response of the conv FS system
figure
plot(t,h,'*');   % Plotting the output
grid on;
yticks([unique(h)]);
title('Self-convolution Fibonacci Sequence','FontSize', 18);
\end{lstlisting}
The sequence $\{1, 2, 5, 10, 20, 38, 71, 130, 235, 420,...\}$ is represented in figure \ref{fig:doubleFS}.

\end{document}